\documentclass[fleqn,usenatbib]{mnras}

\usepackage{newtxtext,newtxmath}

\usepackage[T1]{fontenc}
\usepackage{booktabs}
\usepackage{longtable} 
\usepackage{lipsum} 
\usepackage{array} 
\DeclareRobustCommand{\VAN}[3]{#2}
\let\VANthebibliography\thebibliography
\def\thebibliography{\DeclareRobustCommand{\VAN}[3]{##3}\VANthebibliography}

\usepackage{graphicx}	
\usepackage{amsmath}	

\title[Short title, max. 45 characters]{Luminosity Function of Type II GRBs -- Differences from Long GRBs}

\author[Y. K. Qu et al.]{
Yan-Kun Qu,$^{1,3}$\thanks{E-mail: quyk@qfnu.edu.cn(QYK)}
Zhong-Xiao Man,$^{1}$
 Yu-Peng Yang,$^{1}$
 Shuang-Xi Yi$^{1}$
 and Fa-Yin Wang$^{2}$
\\
$^{1}$School of Physics and Physical Engineering,Qufu Normal University, Qufu,273165, People's Republic of China\\
$^{2}$School of Astronomy and Space Science, Nanjing University, Nanjing 210093, People’s Republic of China
}

\date{Accepted XXX. Received YYY; in original form ZZZ}

\pubyear{\the\year{}}

\begin{document}
\label{firstpage}
\pagerange{\pageref{firstpage}--\pageref{lastpage}}
\maketitle

\begin{abstract}
Gamma-ray bursts (GRBs) are generally categorized into long and short bursts based on their duration ($T_{90}$). Recently, it has been proposed that GRBs  can also be classified into type I (merger) and type II (collapsar) bursts  based on  the different origin.
From a sample of \textit{Swift} long GRBs~(LGRBs) with  a redshift completeness of 60\% and $P \geq 2.6 \, \text{ph} \, \text{cm}^{-2} \, \text{s}^{-1}$, collected through the end of 2023, we identify a pure sample of 146 Type II GRBs.
With this sample, we construct the luminosity function (LF) using both the Broken Power Law (BPL) and Triple Power Law (TPL) models. Our results indicate that, similar to LGRBs, a strong redshift evolution in either luminosity or density is necessary to  accurately account for the observations, regardless of the specific form of the LF assumed. The LF of LGRBs remains a topic of debate, with some studies suggesting it follows a BPL form, while others advocate for a TPL form. In our study, we find that the LF of Type II GRBs tends to favor a BPL model.

\end{abstract}




\section{Introduction}

Gamma-ray bursts~(GRBs) are among the universe's most powerful events, with energy releases up to  $\sim 10^{54}$ erg \citep{Mészáros_2006,annurev:/content/journals/10.1146/annurev.astro.46.060407.145147,KUMAR20151}. 
They are generally categorized into two types based on their duration \citep{1993ApJ...413..281B,2008A&A...484..293Z}, with a boundary set at  $T_{90} = 2s$: short GRBs~(SGRBs) are thought to originate from the merger of compact objects \citep{2013ApJ...769...56F,2013Natur.500..547T}, while long GRBs~(LGRBs) are associated with the collapse of massive stars \citep{1993ApJ...405..273W,2010MNRAS.405...57S}. If this theory holds, the event rate of LGRBs should track the star formation rate~(SFR) \citep{1998ApJ...498..106M,2006ApJ...651..142H}, although the exact relationship between the LGRB event rate and the SFR remains debated ~\citep{2011MNRAS.417.3025V,2012A&A...539A.113E,2015ApJS..218...13Y,2015ApJ...806...44P,WANG20151,2016A&A...587A..40P,Tsvetkova_2017,2019MNRAS.488.5823L}.


\cite{2009ApJ...703.1696Z} proposed a set of criteria to distinguish merger-origin (Type I) GRBs from collapse-origin (Type II) GRBs. Notably, three out of the five Type I GRBs in the gold sample have $T_{90} > 2 s$, which would traditionally classify them as LGRBs. A particularly striking sample is GRB 211211A \citep{2022Natur.612..223R}, which was linked to a kilonova with $T_{90} > 30 s$  and $z = 0.076$. Similarly, GRB 230307A \citep{2024Natur.626..737L}, with $z = 0.0646$ and $T_{90} \sim 35 s$, was identified as originating from a binary neutron star merger (For a more detailed discussion see \citet{2025JHEAp..45..325Z}).

\vspace{-0.25em}

Recently,~\cite{2024ApJ...963L..12P} found that the event rate of LGRBs can be well-explained by a combination of the SFR and the merger rate of compact object binaries.~\cite{2022MNRAS.513.1078D}  discovered that the event rate of high-luminosity LGRBs follows the SFR closely, while the event rate of low-luminosity LGRBs significantly deviates from the SFR.
It is important to note that there is considerable overlap between low luminosity and low redshift sources. The findings of \cite{2024ApJ...963L..12P} and \cite{2022MNRAS.513.1078D} could be seen as complementary perspectives on the same underlying phenomenon.
In contrast to supernova shock breakouts (SBOs) and tidal disruption events (TDEs), whose LFs tend to follow a single power law, \cite{2015ApJ...812...33S} found that LGRBs are better described by a triple power-law (TPL) LF, which is further confirmed by \cite{2021MNRAS.508...52L}.Interestingly, \citet{2021ApJ...914L..40D}  found that the LF of short GRBs also follow a TPL model.

In this Letter, we investigate the LF of Type II GRBs selected from a  LGRBs sample  with  a redshift completeness of $60\%$ and a peak flux of $P \geq 2.6 \, \text{ph} \, \text{cm}^{-2} \, \text{s}^{-1}$. We focus on the differences between the LFs of Type II GRBs and LGRBs. In Section \ref{sec:Samples and Methods}, we outline the criteria for data selection and the process of constructing the LF. Section \ref{Conclusions} presents our results, followed by a brief discussion in Section \ref{Discussion}.

\section{Samples and Methods }

\label{sec:Samples and Methods}
\subsection{Samples}

For constructing GRBs LFs, we revisit the criteria established by \cite{2012ApJ...749...68S} and place emphasis on achieving a balance between redshift completeness and sample size. To evaluate our selection criteria, we introduce a metric \( F = N \times C^3 \), where \( N \) is the ratio of the filtered samples to the total number of \textit{Swift} bursts, and \( C \) represents the redshift completeness after filtering. We set a peak flux threshold of \( P \geq 2.6 \, \text{ph} \, \text{cm}^{-2} \, \text{s}^{-1} \) and include \textit{Swift}/XRT error, Galactic extinction (\( A_V \)), GRB declination (\( \delta \)), and Sun-to-field distance (\( \theta_{\text{Sun}} \)) as free variables in our optimization to identify the conditions that maximize the corresponding \( F \)-value.

Our refined selection yields a sample of 280 GRBs with a redshift completeness of $60\%$. Compared to the sample obtained directly through the original Salvaterra criterion,our method results in an $8\%$ decrease in redshift completeness but a significant $75\%$ increase in sample size, indicating a favorable trade-off between completeness and dataset size. This expanded sample allows us to further refine our selection, enabling us to isolate a pure Type II GRB sample for subsequent analysis.

\textbf{(1) $T_{90,i} \geq 2 s$ }

$T_{90,i}=T_{90}/(1+z)$ represents the intrinsic duration of GRB accounting for the cosmological time dilation effect. The bimodal distribution of GRBs, segmented into long and short categories based on $T_{90}$, was first identified through observations by Konus-Wind \citep{1981Ap&SS..80....3M}, which was confirmed by BATSE~\citep{1993ApJ...413L.101K} and \textit{Swift}~\citep{2013ApJ...764..179B}.~However, relying solely on $T_{90}$ as a classification criterion can sometimes be misleading. For instance, GRB 080913 ($z \sim 6.7$, $T_{90} = 8 \pm 1s$ ) \citep{2009ApJ...693.1610G}, and GRB 090423 ($z \sim 8.2$, $T_{90} = 10.3 \pm 1.1s$ ) \citep{2009Natur.461.1254T} both exhibit intrinsically shorter durations when considering the effect of cosmological time dilation.  Therefore, the intrinsic duration $T_{90,i}$ offers a more reliable metric for distinguishing and analyzing these events.

\textbf{(2) $\epsilon >0.03 $ }

The classification of GRBs based on \(T_{90} = 2 \text{s}\), distinguishing long from short bursts, has been challenged by GRB 060614 \citep{2006Natur.444.1050D}. GRB 060614 has a \(T_{90} \sim 100\text{s}\), which should be classified as a typical long burst. However, its light curve is characterized by a short, hard spike followed by a series of softer gamma-ray pulses. \cite{2007ApJ...655L..25Z} proposed classifying GRBs into two types based on their progenitors: Type I (merger events) and Type II (collapsar events). \cite{2009ApJ...703.1696Z} outlined several criteria for distinguishing between Type I and Type II GRBs, including: (1) \(T_{90i}\), (2) Supernova (SN) association, (3) Host galaxy type, (4) Offset of the GRB from the host galaxy, (5) Surrounding medium type and so on. However, not every GRB has available information for all of these criteria. To maximize the number of GRBs that can be classified, we adopted the parameter \( \epsilon \approx 0.03 \), proposed by \cite{2010ApJ...725.1965L}, as a discriminator between Type I and Type II GRBs.The parameter \( \epsilon \) is defined as:

\begin{equation}
\epsilon = \frac{E_{\gamma,\text{iso},52}}{(E_{p,z,2})^{5/3}},
\end{equation}
where \( E_{\gamma,\text{iso},52} \) represents the isotropic gamma-ray energy in units of \( 10^{52} \, \text{erg} \), and \( E_{p,z,2} \) denotes the spectral peak energy in the cosmic rest frame in units of $100 keV$. 
By applying these selection criteria, we obtained a final sample of 146 GRBs, and used the sample to derive the LF.

\subsection{ANALYSIS METHODs}
The origin of Type II GRBs is linked to the collapse of massive stars. The occurrence of each GRB signifies the death of a short-lived massive star. Consequently, the GRB event rate should be correlated with the SFR as:

\begin{equation}
\psi(z) = \eta \psi_\star(z),
\label{psi}
\end{equation}
where $\psi_{\star}(z)$ represents the SFR in units of $M_{\odot} \, \text{yr}^{-1} \, \text{Mpc}^{-3}$, which can be written as \citep{2006ApJ...651..142H,2008MNRAS.388.1487L},

\begin{equation}
 \psi_{\star}(z) = \frac{0.0157 + 0.118z}{1 + (z/3.23)^{4.66}},
\end{equation}
where $\eta$ denotes the GRB formation efficiency, representing the number of GRBs generated per unit solar mass formed with units of $M_\odot^{-1}$.

To characterize the GRB LF \( \phi(L, z) \), we utilize a general expression derived from a BPL model. In line with the approach by \cite{2021MNRAS.508...52L} , we also propose  a TPL model.

When constructing the LF for LGRBs, three evolution models are commonly discussed: the no evolution model, the luminosity evolution model, and the density evolution model. The no evolution model assumes that the Type II GRB event rate strictly follows the SFR, and the LF does not evolve with redshift, i.e., $L_c(z) = L_{c,0} = constant$. However, both parametric \citep{2012ApJ...749...68S,2019MNRAS.488.4607L,2021MNRAS.508...52L} and non-parametric analyses \citep{2002ApJ...574..554L,2015ApJS..218...13Y,2015ApJ...806...44P,2016A&A...587A..40P} have demonstrated that the no evolution model cannot adequately fit the data, necessitating the inclusion of an evolutionary factor. Therefore, in this study, we focus on the luminosity evolution and density evolution models.

In the density evolution model, the GRB formation rate follows the SFR with an additional redshift-dependent term:
$
\psi(z) = \eta \psi_\star(z)(1 + z)^\delta$, and the LF remains non-evolving. In contrast, the luminosity evolution model assumes that the GRB formation rate is still proportional to the SFR, but the break luminosity evolves with redshift:
$L_{ci}(z) = L_{ci,0}(1 + z)^\delta
$

We employed the maximum likelihood estimation technique to refine the free parameters of our model \citep{1983ApJ26935M,2019MNRAS.490..758Q,2025ApJ...982..148Q}. The likelihood function is given as:

\begin{equation}
    \mathcal{L}=\mathrm{e}^{(-N_{exp})} \prod_{i=1}^{N_{obs}} \Phi(L_{i},z_{i},t_{i}) .
    \label{eq:L}
\end{equation} 

In this framework, $N_{exp}$ denotes the expected number of GRBs detected, and $N_{obs}$ represents the actual observed sample size. The rate of GRBs per unit time is given by $\Phi(L, z, t)$, where redshift ranges from $z$ to $z + dz$, and luminosity spans from $L$ to $L + dL$. The specific functional form of $\Phi(L, z, t)$ is defined as:

\begin{equation}
\begin{split}
    \Phi(L, z, t) &= \frac{d^{3} N}{dtdzdL} 
    = \frac{d^{3}N}{dtdVdL} \times \frac{dV}{dz} \\
    &= \frac{\Delta \Omega}{4\pi}  \theta(P)  \frac{\psi(z)}{(1+z)}  \phi(L,z) \times \frac{dV}{dz},
\end{split}
\label{eq:Phi}
\end{equation}
where $\Delta \Omega = 1.33 \, \text{sr}$ represents the field of view of \textit{Swift}/BAT, $\psi(z)$ represents the comoving event rate of GRBs (Eq. \ref{psi}), $\phi(L,z)$ represents the normalized GRB LF.

The detection efficiency, $\theta(P)$, is decomposed as $\theta(P) \equiv \theta_{\gamma}(P) \theta_{z}(P)$, where $\theta_{\gamma}(P)$ is the probability of detecting a burst with peak flux $P$. Given that our sample is constrained by $P \geq 2.6 \, \text{ph} \, \text{cm}^{-2} \, \text{s}^{-1}$, we assume $\theta_{\gamma}(P)=1$, rendering the selection effect negligible. The term $\theta_{z}(P)$ describes the probability of measuring the redshift for a given $P$. In prior works, for high redshift-complete samples \citep{2012ApJ...749...68S,2016A&A...587A..40P}, this term has been approximated as unity \citep{2019MNRAS.488.4607L}, due to redshift completeness exceeding $80\%$. However, for our dataset with a redshift completeness of $60\%$, we model the relationship between $\theta_{z}(P)$ and $P$ by fitting an empirical function: 
\begin{equation}
\theta_{z}(P) = \frac{1}{1 + (1.28 \pm 0.21) \times (0.95 \pm 0.01)^P}.
\end{equation}

The anticipated number of GRBs can be expressed as:

\begin{equation}
\begin{split}
    N_{\text{exp}} = \frac{\Delta \Omega T}{4 \mathrm{\pi}} \int_{z_{\text{min}}}^{z_{\text{max}}} \int_{\max[L_{\text{min}}, L_{\lim}(z)]}^{L_{\text{max}}} \theta(P(L, z)) \frac{\psi(z)}{1+z} \\ 
    \times \phi(L, z) \, dL \, dV(z),
\end{split}
\label{eq:Nexp}
\end{equation}
where $T \sim 19$ years reflects the duration of \textit{Swift}'s operation for this dataset.  The luminosity threshold in Eq.~\ref{eq:Nexp} is given by:

\begin{equation}
    L_{\lim}(z) = 4 \pi D_{L}^{2}(z) P_{\lim} \frac{\int_{1/(1+z) \, \text{keV}}^{10^4/(1+z) \, \text{keV}} E N(E) \, dE}{\int_{15 \, \text{keV}}^{150 \, \text{keV}} N(E) \, dE},
\end{equation}
where $P_{\lim} = 2.6 \, \text{ph} \, \text{cm}^{-2} \, \text{s}^{-1}$ represents the sample’s peak flux threshold. 
\begin{table*}
\centering
\caption{The best fit parameters of each model. }
\begin{tabular}{cccccccccc}
\toprule
Model & Evolution parameter &$\eta$ &  $a$ & $b$ &$c$& log $L_c$ (erg s$^{-1}$) & ln $\mathcal{L}$ & AIC&BIC \\
 &  &  ($10^{-8} M^{-1} _{\odot}$) &  & & &   & & \\ \midrule
  \multicolumn{10}{c}{\textbf{Broken power-law luminosity function}} \\
 \midrule

Luminosity evolution & $\delta = 1.65^{+0.20}_{-0.22}$ &$3.89^{+0.88}_{-0.86}$ & $-0.71^{+0.04}_{-0.04}$ & $-1.98^{+0.20}_{-0.19}$ & & $52.82^{+0.12}_{-0.13}$ & -121.49 & 252.98&267.90 \\
Density evolution & $\delta = 1.40^{+0.22}_{-0.21}$ &$ 4.11 ^{+0.93}_{-0.94}$&$-0.66^{+0.04}_{-0.04}$ & $-1.26^{+0.21}_{-0.19}$ & &$52.93^{+0.10}_{-0.11}$ & -121.09 & 252.18&267.10 \\ \midrule
  \multicolumn{10}{c}{\textbf{Triple power-law luminosity function}} \\
\midrule
Luminosity evolution & $\delta = 1.63^{+0.23}_{-0.25}$ &$9.25^{+0.91}_{-0.90}$ & $-1.43^{+0.25}_{-0.25}$ & $-0.68^{+0.05}_{-0.05}$ & $-1.96^{+0.26}_{-0.25}$ &$49.77^{+0.10}_{-0.11}$ $52.76^{+0.12}_{-0.13}$ & -121.01 & 256.02&276.91 \\
Density evolution & $\delta = 1.46^{+0.22}_{-0.21}$ &$ 8.23 ^{+0.88}_{-0.85}$&$-1.61^{+0.23}_{-0.23}$ & $-0.57^{+0.07}_{-0.07}$ &$-1.10^{+0.11}_{-0.10}$ &$50.22^{+0.12}_{-0.13}$ $52.61^{+0.13}_{-0.14}$ & -120.15 & 254.30&275.19 \\

\bottomrule
\end{tabular}
\begin{minipage}{\textwidth}
\footnotesize
\textbf{Note.} The parameter values are determined as the medians of the best-fitting parameters from the Monte Carlo sample. The errors represent the $68\%$ containment regions around the median values.
\end{minipage}
\label{table1}
\end{table*}

\begin{figure*}
    \centering
    \includegraphics[width=0.9\textwidth]{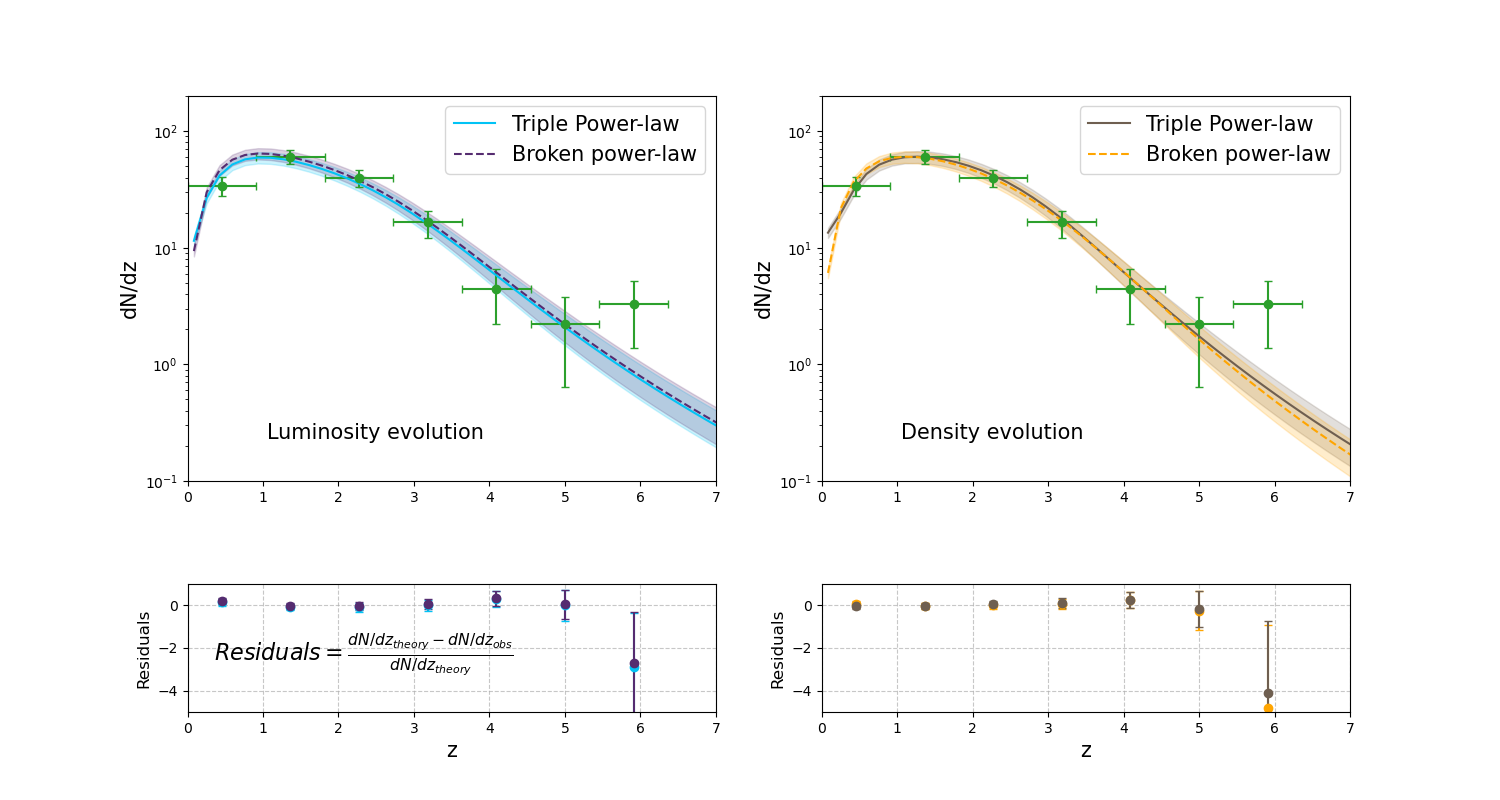}
    \caption{\textbf{Upper Panel:}Green data points represent the redshift distribution of 146 type II GRBs with $P \geq 2.6 \, \text{ph} \, \text{cm}^{-2} \, \text{s}^{-1}$ in our sample. The solid and dashed lines  correspond to the expected distributions from different best-fitting models. The shadow regions represent the 1$\sigma$ confidence region of the corresponding models.\textbf{Lower panel:} The residuals corresponding to the upper panel, with colors matching those in the upper panel.}
    \label{fig:dNdz}
\end{figure*}

\begin{figure*}
    \centering
    \includegraphics[width=0.8\textwidth]{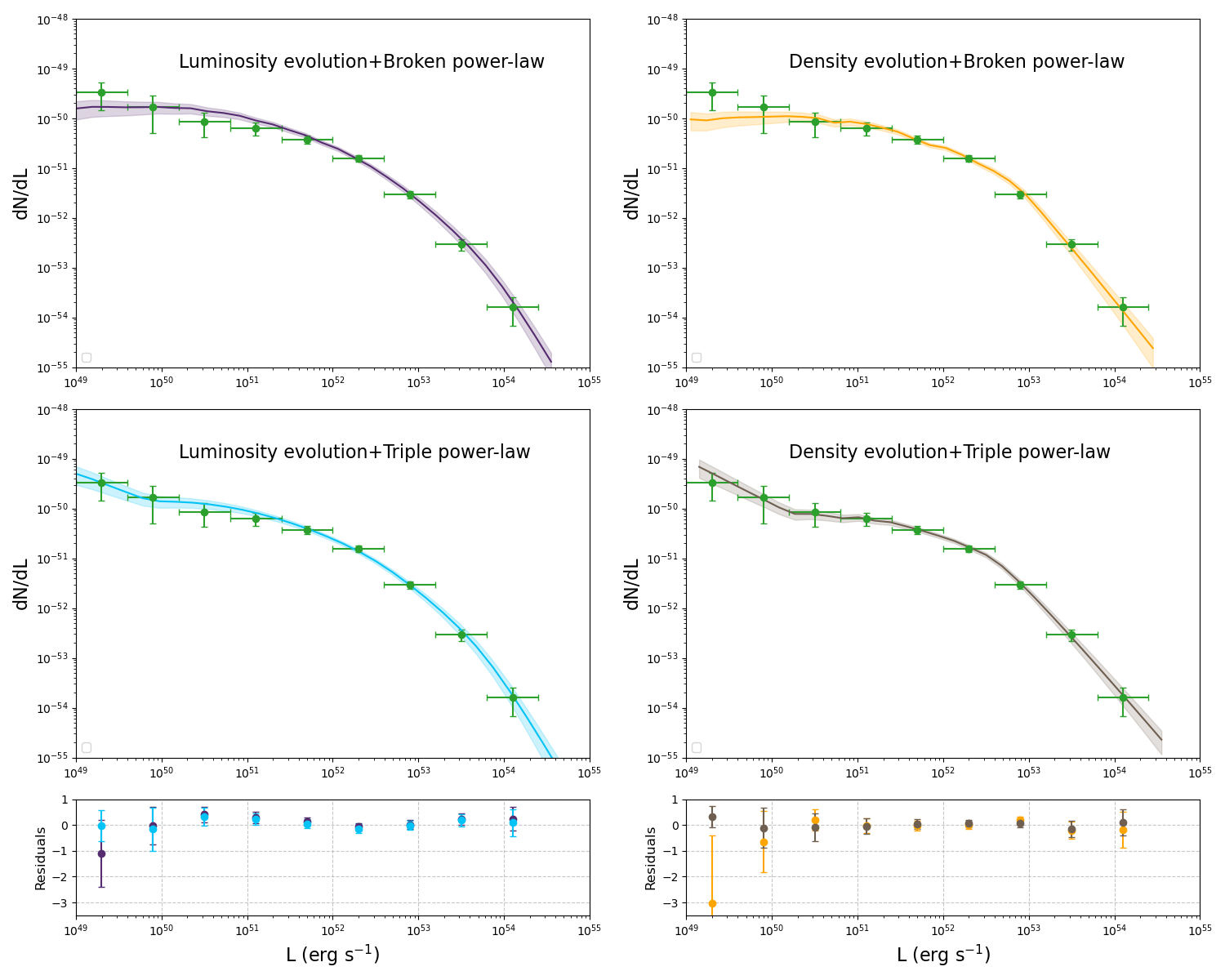}
    \caption{Green data points represent the luminosity distribution of 146 type II GRBs with $P \geq 2.6 \, \text{ph} \, \text{cm}^{-2} \, \text{s}^{-1}$ in our sample. The solid lines  correspond to the expected distributions from different best-fitting models. Shadow regions represent the 1$\sigma$ confidence region of the corresponding models.
    The bottom row displays the residuals corresponding to the data shown in the top two rows, with colors matching those in the top two rows.}
    \label{fig:dNdL}
\end{figure*}

\section{RESULTS}
We employed the Markov Chain Monte Carlo (MCMC) method to derive the best-fit parameters  and corresponding maximum likelihood for our models. The chi-square statistic:
\begin{equation}
\begin{aligned}
\chi^2 &=  - 2\ln \mathcal{L}. 
\end{aligned}
\end{equation}

Considering the sample size and model parameters, we then calculated the reduced chi-square for each model. For the Luminosity Evolution Model, the reduced chi-square for the BPL model is 1.72, and for the TPL model, it is 1.74. For the Density Evolution Model, the reduced chi-square for the BPL model is 1.72, and for the TPL model, it is 1.73. From the perspective of reduced chi-square, the BPL model performs slightly better than the TPL, but the difference is not substantial. Therefore, the inclusion of Akaike Information Criterio (AIC) \citep{1974ITAC...19..716A} and Bayesian Information Criterion (BIC) \citep{24ce203a-855a-3aa9-952f-976d23b28943} , which impose more significant penalties for model complexity, is crucial for the analysis.
 The corresponding definitions are as follows:
\begin{equation}
\begin{aligned}
\text{AIC} &= 2k - 2\ln \mathcal{L}, \\
\text{BIC} &= k \log N - 2\ln \mathcal{L},
\end{aligned}
\end{equation}
where $k$ is the number of model parameters (degrees of freedom), $\mathcal{L} $ is the maximum likelihood, and $N$ is the total number of data points. The results are summarized in Table \ref{table1}.
The relative probabilities of the models are computed using the following formula \citep{Burnham_04,2007MNRAS.377L..74L}:
\begin{equation}
\begin{split}
    P(M_{i})=\frac{\mathrm{e}^{-AIC_{i}/2}}{\mathrm{e}^{-AIC_{1}/2} + \mathrm{e}^{-AIC_{2}/2}}.
\end{split}
\label{eq:aIC}
\end{equation}

For luminosity evolution, based on the AIC, the probability of the TPL model being correct relative to the BPL model is $17.9\%$, while the probability of the BPL model being correct relative to the TPL model is $82.1\%$. When considering the BIC, which imposes a stronger penalty for model complexity, the probability of the TPL model being correct relative to the BPL model decreases to $1.09\%$, whereas the probability of the BPL model being correct relative to the TPL model increases to $98.91\%$.
For density evolution, based on the AIC, the probability of the TPL model being correct relative to the BPL model is $25.8\%$, while the probability of the BPL model being correct relative to the TPL model is $74.2\%$. When evaluated using the BIC, the probability of the TPL model being correct relative to the BPL model is only $1.73\%$, while the probability of the BPL model being correct relative to the TPL model rises to $98.27\%$.

Additionally, we also calculated the relative differences $\Delta \text{AIC (BIC)} $ between the models. According to information theory standards \citep{2020MNRAS.494.2576C,2022A&A...661A..71H}:

(i) $\Delta \text{AIC (BIC)} \in [0, 2]$~indicates weak evidence for the reference model, with difficulty distinguishing between the models;
 
(ii) $\Delta \text{AIC (BIC)} \in (2, 6]
$~indicates mild evidence against the given model relative to the reference model;

(iii) $\Delta \text{AIC (BIC)} >6
$~indicates strong evidence against the given model, suggesting it should be rejected.

In our analysis, for both luminosity and density evolution, the BIC value for the TPL model is more than 6 greater than that of the BPL model, providing strong evidence to reject the TPL model.

In Figure \ref{fig:dNdz}, we present the redshift distribution of the 146 Type II GRBs from our sample. The upper left (right) panel illustrates the redshift distribution under
the luminosity (density) evolution with a BPL and TPL LF.  The lower panel displays the residuals corresponding to the upper panel, which is defined as
\begin{equation}
\text{Residuals} = \frac{dN/dz_{\text{theory}} - dN/dz_{\text{obs}}}{dN/dz_{\text{theory}}}.
\end{equation}
Notably, all four models provide a good fit to the observed redshift distribution.
Figure \ref{fig:dNdL} presents the luminosity distributions corresponding to the four models. While the low luminosity region of  Figure \ref{fig:dNdL} suggests that the TPL model provides a better fit, the TPL model has two additional degrees of freedom compared to the BPL model, meaning that its maximum likelihood value $\ln \mathcal{L}$ cannot be lower than that of the BPL model. Furthermore, it is important to note that our approach is not to directly fit the luminosity distribution and redshift distribution, but rather to use the maximum likelihood method outlined in Equation \ref{eq:L} of this paper. In this method, each GRB sample is given equal weight, and there are very few low-luminosity samples. Therefore, even though there may be a visible difference between the models in the residuals at low luminosity, the limited number of low-luminosity samples has minimal impact on the final fitting result. Considering this, it is necessary to use model comparison tools that account for model complexity, such as AIC and BIC.


By comparing the AIC and BIC of the models, we conclude that, compared to the TPL model, the LF of Type II GRBs is more consistent with a BPL. Interestingly, the form of the LF for LGRBs remains a subject of ongoing debate. Different studies have employed varying sample selection criteria and analytical methods, leading to differing conclusions. Some studies suggest that the LF of LGRBs favors the BPL \citep{2010MNRAS.406.1944W, 2015ApJ...806...44P,2023ApJ...958...37D}, while others argue that it follows a TPL\citep{2021MNRAS.508...52L}.

Our results indicate that the goodness of fit for both evolution models is very similar, making it challenging to distinguish between the two. This is consistent with previous findings using LGRBs to construct LFs \citep{2012ApJ...749...68S,2021MNRAS.508...52L}. Regardless of the model, both suggest that a strong evolution term is necessary to better fit the data. Utilizing a BPL LF, our analysis reveals significant luminosity evolution with \( \delta = 1.65^{+0.27}_{-0.31} \). This result is consistent with earlier parametric studies (e.g., \cite{2012ApJ...749...68S}, \cite{2019MNRAS.488.4607L}) and broadly aligns with non-parametric investigations such as \cite{2002ApJ...574..554L} (\( \delta = 1.4 \pm 0.5 \)) and \cite{2015ApJ...806...44P} ( \( \delta = 2.3 \pm 0.8 \)).
For the density evolution model, our results indicate that an evolution term with $\delta = 1.4$ provides a better fit to the data. Several previous studies have reached similar conclusions, such as \cite{2008ApJ...673L.119K}, \cite{2013A&A...556A..90W} and \cite{2021MNRAS.508...52L}. One commonly proposed explanation for this is related to the metallicity and rotation rates of stars at higher redshifts. Specifically, for stars with masses $M \geq 30 M_{\odot}$, GRBs can be produced via the conventional collapse model \citep{1993ApJ...405..273W}. For stars with $12 M_{\odot} < M < 30 M_{\odot}$, GRB production is more closely related to metallicity. Low metallicity and rapid rotation are thought to facilitate the efficient production of GRBs (see \cite{2013A&A...556A..90W} for details). In addition, alternative explanations have been put forward, including cosmic metallicity evolution \citep{2008MNRAS.388.1487L,2006ApJ...638L..63L,Wang2022}, cosmic strings \citep{2010PhRvL.104x1102C}, and evolving initial mass function of stars 
 \citep{2011ApJ...727L..34W}.

\label{Conclusions}
\section{Conclusions and Discussion}

\label{Discussion}
From a sample of GRBs with a redshift completeness of $60\%$ and a peak flux $P \geq 2.6 \, \text{ph} \, \text{cm}^{-2} \, \text{s}^{-1}$, recorded up to the end of December 2023, we selected 146 Type II GRBs.
~We constructed the LFs using both the BPL and TPL models, considering two scenarios: luminosity evolution and density evolution. Similar to the findings for LGRBs, our results indicate that the goodness of fit for luminosity evolution and density evolution is comparable, making it difficult to distinguish between the two evolution models. Both models suggest that a strong evolutionary term is required for a better fit to the data. The LF of LGRBs remains a topic of debate, with some studies suggesting it follows a BPL form, while others advocate for a TPL form. We found that   Type II GRBs show a preference for the BPL model. Moreover, the results of both the AIC and BIC strongly suggest that the TPL model can be nearly ruled out.


\cite{2022MNRAS.516.2575J} attempted to constrain cosmological parameters using the Amati relation. Their study revealed that when fitting the $ E_{p}-E_{iso}$ relation of GRBs at different redshifts, the slope is largest at low redshift and gradually decreases with increasing redshift, stabilizing around at redshift $z\sim 2$. The contamination of LGRB samples by  merger GRBs could provide a kind of explanation for this phenomenon. Overall, for the same $E_{p}$ value, merger GRBs exhibit lower $E_{iso}$ values compared to collapsar GRBs. At low redshift, merger GRBs constitute a substantial portion of LGRBs, resulting in a steeper $ E_{p}-E_{iso}$ slope. Unlike LGRBs, whose event rates follow the SFR, merger GRBs are expected to follow the delayed SFR \citep{2018MNRAS.477.4275P}. As redshift increases, the proportion of  merger GRBs within LGRBs diminishes, becoming negligible around redshift $z\sim 2$.

This contamination of LGRB samples by merger GRBs can also explain the findings of \cite{2023ApJ...958...37D}, where they  found that the event rate of high-luminosity LGRBs closely tracks the SFR, while the event rate of low-luminosity LGRBs deviates from it. In general, merger GRBs have lower luminosities than collapsar GRBs. Therefore, the high-luminosity sample is predominantly composed of collapsar GRBs, which explains its close alignment with the SFR. The low-luminosity sample, however, includes a mixture of merger GRBs, resulting in deviations from the SFR.

Classifying GRBs into two categories based on $T_{90} = 2 \, \mathrm{s} $ is a rather qualitative approach. This classification has been challenged by recent observations, such as a SGRB with an associated supernova (GRB 200826A\citep{2021NatAs...5..917A}) and two LGRBs associated with kilonovae (GRB 211211A\citep{2022Natur.612..232Y} and GRB 230307A\citep{2024Natur.626..737L}). These cases highlight the limitations of using \( T_{90} \) alone to distinguish GRBs of different physical origins.
\cite{2024ApJ...963L..12P} demonstrated that the event rate of LGRBs follows the SFR well at redshifts $z > 2$ , but significantly exceeds the SFR at $z < 2$ .  By subtracting the SFR from the LGRB event rate, they found that the residual component closely resembles the shape of a delayed star formation rate\citep{2018MNRAS.473.3385P}, suggesting that up to $60\%$ of LGRBs at $z < 2$ may not originate from collapsars. Building on this hypothesis, \citet{2024ApJ...976..170Q,2025MNRAS.540L...6Q} assumed that all LGRBs at $z > 2$  are collapsars, and used the high-redshift LGRBs to constrain their luminosity function, which was then extrapolated to lower redshifts. The extrapolated luminosity function for high-redshift LGRBs was found to significantly underpredict the observed event rate at $z < 2$ , further supporting the conclusion that a substantial fraction of low-redshift LGRBs are not collapsars.
Notably, GRBs at redshifts lower than 2 comprise approximately $50\%$ of all GRBs with measured redshifts. Thus, efficiently distinguishing between merger and  collapsar samples within LGRBs becomes crucial.

\begin{table*}
\centering
\caption{List of 146 GRBs meeting our selection criteria.\label{table2}}

\begin{tabular}{ l l l l l l l l  }
\hline
\hline

GRB(Ref) & GRB(Ref) & GRB(Ref) & GRB(Ref) & GRB(Ref) & GRB(Ref) & GRB(Ref) & GRB(Ref)   \\
\hline
050219A(1, 3) & 050318(1, 4) & 050525A(2, 3) & 050603(3) & 050802(1, 4) & 051008(1, 4) &
051109A(13) & 051111(1, 3) \\ 060210(15) & 060306(1, 4) & 060418(1, 4) & 060505(1) & 060814(1, 4) & 060908(1, 4) & 060912A(2, 3) & 060927(13) \\ 061007(1, 4) & 061121(1, 3) &
061126(1, 3) & 061222A(2, 3) & 070306(1, 4) & 070328(1, 4) & 070521(15) & 071112C(1, 3) \\
071117(1, 3) & 080319B(1, 3) & 080319C(1, 3)&
080411(1, 4) & 080413A(2, 3) & 080413B(1, 3) &
080430(1, 4) & 080602(1, 4) \\ 080603B(14) & 080605(1, 4) & 080607(14) & 080721(14) &
080804(15) & 080916A(2, 3) & 081007(1, 4) & 081121(14) \\ 081203A(2, 3) & 081221(15) &
081222(14) & 090201(1, 3) & 090424(1, 4) & 090709A(3) & 090715B(16) & 090812(16) \\
090926B(1, 3) & 091018(1, 4) & 091020(1, 4) & 091127(1, 4) & 091208B(1, 3) & 100615A(1, 3) &
100621A(1, 3) & 100704A(2, 3) \\ 100728A(1, 3) & 100728B(1, 3) & 110205A(1, 3) & 110422A(1, 3) &
110709B (1, 3) & 110731A(1, 3) & 111008A(1, 3) & 111228A (1, 3) \\ 120119A(1, 3) & 120326A(1, 3) &
120729A(1, 3) & 120802A(1, 3) & 120811C(15) & 120907A(1, 3) & 121209A(1, 3) & 130420A(1, 3) \\
130427A(1, 3) & 130427B(1, 3) & 130505A(1, 4) & 130514A(1, 3) & 130606A(1, 3) & 130701A(1, 3) &
130831A(1, 3) & 130907A(1, 3) \\ 130925A(2, 3) & 131030A(1, 3) & 131105A(1, 3) & 140206A(16, 17) &
140213A(1, 3) & 140419A(1, 3) & 140506A(1, 3) & 140512A(1, 3) \\ 140629A(1, 3) & 140703A(1, 3) &
141220A(1, 3) & 141221A(1, 7) & 150206A(1, 4) & 150301B(1, 4) & 150314A(1, 4) & 150323A(1, 4) \\
150403A(1, 4) & 151021A(1, 5) & 151027A(1, 3) & 160131A(1, 4) & 160410A(2, 4) & 160425A(1) &
160804A(1, 21) & 161014A(1, 4) \\ 161017A(1, 6) & 161117A(1, 18) & 161129A(1, 4) & 170202A(1, 7) &
170604A(1, 4) & 170705A(1, 8) & 170903A(1) & 180205A(1, 22) \\ 180314A(2, 19) & 180325A(1, 9) &
180620B(1, 23) & 180720B(1, 2) & 181020A(1, 10) &
181110A(1, 24) & 190106A(1, 25) & 190114C(1, 26) \\
190324A(2, 20) & 190719C(1, 11) & 191004B(2, 12) & 191019A(2) & 191221B(1, 2) & 200528A(1, 2) &
200829A(1, 2) & 201024A(1, 2) \\ 201104B(1, 2) & 201216C(1, 2) & 210104A(1, 2) & 210112A(1, 2) &
210207B(1, 2) & 210210A(1, 2) & 210217A(1, 2) & 210411C(1, 2) \\
210610B(1, 2) & 210619B(1, 2) &
210822A(1, 2) & 211211A(1, 2) & 211227A(1, 2) & 220101A(1, 2) & 220521A(1, 2) & 230325A(1, 2) \\
230818A(1, 2) & 231111A(1, 2) &  &  \\
															
\hline
\end{tabular}

\begin{minipage}{\textwidth}
\footnotesize
\noindent \textbf{Note:} 146 \textit{Swift} GRB used in this study.The spectra and redshift references are from the following sources: 
(1) \url{https://www.mpe.mpg.de/~jcg/grbgen.html};
(2) \url{https://swift.gsfc.nasa.gov/archive/grbtable/};
 (3) \cite{2020ApJ...893...77W};
 (4) \cite{2007ApJ...671..656B};
 (5) \cite{2015GCN.18433....1G}
 (6) \cite{2016GCN.20082....1F}
 (7) \cite{2017GCN.20604....1F}
 (8) \cite{2017GCN.21297....1B}
 (9) \cite{2018GCN.22546....1F}
 (10) \cite{2018GCN.23363....1T}
 (11) \cite{2019GCN.25130....1P}
 (12) \cite{2019GCN.25974....1S}
 (13) \cite{2008MNRAS.391..577A}
 (14) \cite{2009A&A...508..173A}
 (15) \cite{2019MNRAS.486L..46A}
 (16) \cite{2016A&A...585A..68W}
 (17) \cite{2020ApJ...893...46V}
 (18) \cite{2016GCN.20192....1M}
 (19) \cite{2018GCN.22513....1T}
 (20) \cite{2019GCN.24002....1H}
 (21) \cite{2016GCN.19769....1B}
 (22) \cite{2018GCN.22386....1V}
 (23) \cite{2018GCN.22813....1M}
 (24) \cite{2018GCN.23424....1F}
 (25) \cite{2019GCN.23637....1T}
 (26) \cite{2019GCN.23707....1H}
 \end{minipage}

\end{table*}

\section*{Acknowledgments}

We thank the anonymous referee for insightful comments and constructive suggestions, which have significantly improved the quality of this manuscript. We are also grateful to Zeng Hou-dun and Lan Guang-xuan for the valuable discussions regarding our results. This work was supported by the National Natural Science Foundation of China
(grant No. 12494575) and the Shandong Provincial Natural Science Foundation (Grant No. ZR2021MA021).

\section*{Data Availability}
All data used in this paper are public.



\bibliographystyle{mnras}
\bibliography{example} 





\bsp	
\label{lastpage}
\end{document}